\documentclass[twocolumn,letterpaper,english]{revtex4}
\usepackage[T1]{fontenc}
\usepackage[latin1]{inputenc}
\usepackage{graphicx}

\makeatletter

\usepackage{graphicx}

\usepackage{babel}

\usepackage{babel}

\def\ra{{$\rightarrow$}\ }
\def\eion{{$(e~+~ion)$}\ }

\def\fexvii{{\rm Fe {\sc xvii}}\ }
\def\fexviii{{\rm Fe {\sc xviii}}\ }
\def\fexix{{\rm Fe {\sc xix}}\ }
\def\en{{$n$\ }}

\newcommand{\be}{\begin{equation}}
\newcommand{\ee}{\end{equation}}
\makeatother
\begin{document}

\title{Large enhancement in high-energy photoionization of \fexvii and
missing continuum plasma opacity}


\author{Sultana N. Nahar$^\ast$ and Anil K. Pradhan$^\dagger$} 

\affiliation{Department of Astronomy,
The Ohio State University, Columbus, Ohio
43210.}

\begin{abstract}

Aimed at solving the outstanding problem of solar opacity, 
and radiation transport plasma models in general, 
we report substantial photoabsorption in the
high-energy regime due to atomic core photo-excitations 
not heretofore considered. In extensive 
R-Matrix calculations of unprecedented complexity for an important iron 
ion \fexvii (Fe$^{16+}$), with a wave function expansion of 99 \fexviii
(Fe$^{17+}$) LS core states from $n\leq 4$ complexes 
(equivalent to 218 fine structure
levels), we find: i) up to orders of magnitude enhancement 
in background photoionization cross 
sections, in addition to strongly peaked photo-excitation-of-core 
resonances not considered in current opacity models, and 
ii) demonstrate convergence with respect to successive core excitations. 
The resulting increase in the monochromatic continuum, and 35\% in
the Rosseland Mean Opacity, are compared with the
"higher-than-predicted" iron opacity measured at 
the Sandia Z-pinch fusion device at solar interior conditions.

 \textbf{~}
\textbf{PACS number(s)} :32.80.Aa,32.80.Fb,32.80.Zb,95.30.Ky
 \end{abstract}
\maketitle

Radiation transport and light-matter interactions depend
fundamentally on plasma opacity, which in turn entails an intricate
interplay of atomic and plasma effects under local conditions. 
In astrophysics, monochromatic
and mean opacities, averaged over photon-particle distributions, are
crucial to the determination of not only emergent spectra but also
the chemical composition of the source. 
Surprisingly, elemental abundances of the Sun have been called into
question in recent years, and revised downwards
by up to $\sim$50\% for common elements such as C, N, O and Ne \cite{asp09}. 
But that has 
posed an outstanding problem for stellar interior and helioseismic models 
\cite{christ09}. 
Recent laboratory measurements 
of monochromatic iron opacity at the Sandia Z-pinch machine of plasma
created at conditions similar to those near the 
convection and radiative zone boundary disagree considerably with current 
opacity models by 30-400\% in the wavelength range of 7-13\AA~ \cite{bail}. 
Discrepancies are evident not only at and between energies where discrete 
transitions occur but, more surprisingly, the measured opacity is 
substantially higher in the high-energy
feature-less bound-free continua (or photoionization) of a few abundant iron 
ions \fexvii, \fexviii and \fexix. 
The {\it upward} revision of opacities \cite{bail} tends to support 
the revised {\it lower} solar abundances. Indeed, the measured Z opacity
for iron alone enhances the solar mixture opacity of all abundant
elements by $\sim$7\%, about
half of the 15\% required to reconcile with stellar interior and
helioseismic models. We show that hitherto neglected atomic and
plasma physics may have a significant bearing on the missing opacity
problem.
 
Current 
opacity models generally consider photoexcitations into the continuum 
as bound-bound 
transitions, and not as resonant transitions into quasi-bound
auotoionizing levels \cite{n11}. However, while
resonances may explain larger broadening of observed features in the Sandia 
experiments, that would not countenance uniformly higher monochromatic 
opacity in the (e~+~ion) bound-free continuum compared to 
theoretical calculations \cite{bail}. 
Atomic codes for opacities such as
the Opacity Project (OP \cite{op}), ATOMIC, OPAS and SCO-RCG (references in
\cite{bail}), typically 
employ a wave function representation with {uncoupled} excitation
channels into the residual core ion. 
Hence the photoionized atom or ion wavefunction expansion is
incomplete, which 
results in (i) 
neglect of interference among overlapping infinite Rydberg series of 
autoionizing resonances converging 
on to excited cores, 
and large photoexcitation-of-core (PEC) resonances with characteristic
asymmetric profiles
(e.g. \cite{aas}), and (ii) ``jumps" that enhance the background cross 
sections at excitation threshold(s) belonging to levels of successively 
higher-$n$ complexes. The present coupled channel calculations show such 
jumps and enhancements in the \eion continua, and
that {\it coupling of the photoionizing 
level and the 
residual ion system requires completeness in terms of converged background up 
to sufficiently high energies.} In this {\it Letter} we report such 
completeness in 
calculations for the coupled \fexvii \ra (e~+~\fexviii) system using the 
Breit-Pauli R-Matrix (BPRM) methodology, and demonstrate the aforementioned 
enhancement up to convergence and relevance to opacity models.
Our study focuses on \fexvii which is the highest 
contributor to opacity in the range 
T $\sim$ 150-200 eV (11-15 Ry or $2 \times 10^{6.24-6.37}$K) and   
electron densities N$_e \sim 10^{21-23}$ cm$^{-3}$, though
the ionization fractions of the three dominant ions \fexvii, \fexviii
and \fexix in local thermodynamic equilibrium (LTE) at the Z temperature 182
eV (13.4 Ry) are 0.19, 0.38 and 0.29 respectively \cite{bail,n11}.
 
Monochromatic plasma opacity $\kappa_\nu$ largely depends on radiation 
absorption through bound-bound (bb) photo-excitation and bound-free (bf) 
photoionization as follows:
\begin{equation}
\kappa^{bb}_{\nu}(i \rightarrow j) = {\pi e^2\over mc} N_i f_{ij} \phi_{\nu};~~
\kappa^{bf}_{\nu} = N_i \sigma_{PI}(\nu)
\end{equation}
where $f_{ij}$ is the oscillator strength, $\sigma_{PI}$ is the photoionization
cross section, $N_i$ is the ion density, and $\phi_{\nu}$ is a profile
factor. While oscillator strengths are discrete, the cross sections form a 
continuum replete with resonant and background features. The contribution of 
$\sigma_{PI}$ to the opacity may be expressed as an effective oscillator 
strength integrated over a given energy range $\Delta E$
\begin{equation}
<f>_{\Delta E} = \sum_k^\infty \int_{\epsilon = E_k}^\infty
\frac{df_{ik}}{d\epsilon} d\epsilon, 
\end{equation}
where $df_{ik}/d\epsilon$ represents the photoionization continua that are 
now discretized according to excitation into ion core levels $k$
from an initial level $i$.
This is a non-trivial extension. In principle, {\it all} 
possible bound levels of the core ion contributing to the total 
(resonant+background) bound-free continuum, or $\sigma_{ik}$ from an initial 
bound level $i$ into a final
\eion continuum with reference to the excited core levels $k$ 
of the residual ion, should be included until convergence and no further
enhancement. 
Explicitly including
successive core excitations in $\sigma_{PI}$,
\be 
<f>_{\Delta E} = \frac{1}{4\pi \alpha a_o^2} \sum_{E_1(k)}^{E_{n'}} \int_{\epsilon \geq
E_k}^\infty \sigma_{ik} d\epsilon_k, 
\ee
where $E_1$ denotes the ground state of the core ion and $E_{n'}$ the
highest level included in the coupled-channel calculations. 
We calculated $\sigma_{ik}$ in the
close coupling (CC) approximation using the R-matrix method \cite{n11}.
We consider all possible dipole
allowed transitions, which are typically the dominant ones, into
states formed by configurations with the outer electron in $n$=2, 3, 4 shells. 

Fig.~1  shows the photo-coupled [\fexvii-\fexviii] 
system with a schematic diagram of the energy level structure of the core 
\fexviii ion included in the current calculations. The OP photoionization 
calculations for \fexvii included only the ground complex of \fexviii, a 2 
LS term expansion $1s^22s^22p^5 \ (^2P^o)$ and $1s^22s2p^6 \ (^1S)$ 
(abbreviated as 2CC) \cite{scott}. The OP and other opacities
calculations include these excited configurations, 
and inter-channel couplings and resonance structures, 
perturbatively. But that, for
example, does not account for asymmetric autoionization profiles or
coupled core excitations affecting background opacity.
In a previous work \cite{n11} we
considered 30 LS terms or 60 coupled fine structure levels up to the $n$ = 3 
complex; we refer to it as 30CC. 
We carried 
out the present calculations with a considerably larger expansion of the 
basis set up to 99 LS terms or 99CC, corresponding to 218 fine structure 
levels up to $n = 4$. 
The size of the Hamiltonian to be diagonalized in R-Matrix calculations
increases as the total number of channels, hence the present
calculations
are more than an order of magnitude larger than in \cite{n11}.
 \begin{figure}
 \begin{center}
\includegraphics[width=85mm,height=55mm]{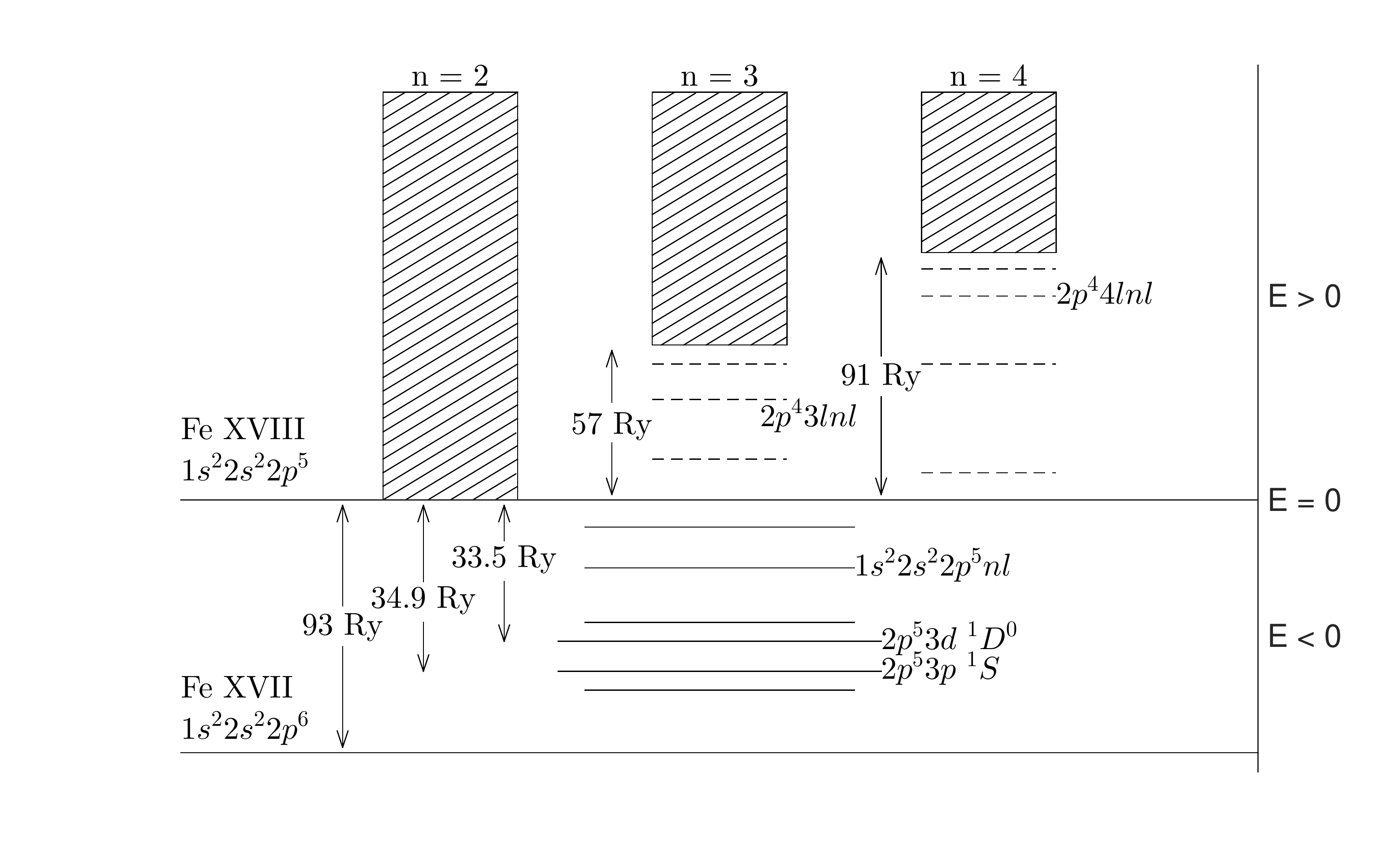}
 \end{center}
\vskip -0.35in
 \caption{Schematic representation of (e~+~\fexviii) $\rightarrow$ \fexvii 
bound and autoionizing levels included in the coupled channel calculations. 
The three results compared herein refer to calculations including 2 LS 
terms of $n = 2$ complex under the OP, previous 30 LS terms (60 fine 
structure levels) of $n = 2,3$ \cite{n11}, and the present 99 LS terms 
with $n = 2,3,4$ in the total wave function expansion. Illustrative
results presented for photoionization in Fig.~3
of the 2 excited bound \fexvii levels 
are highlighted (longer solid lines):  $2p^53p \ ^1S$ and $2p^53d \ ^1D^o$.
\ 
 \label{fig:f1}}
 \end{figure}
We calculated photoionization cross sections for all 283 bound LS states of 
\fexvii up to $n = 10$ and $l$ = 9, corresponding to 454 fine structure
levels obtained in the previous work \cite{n11}. The OP calculation 
\cite{scott} has a much smaller number, 181 bound LS terms.
Without loss of generality, and to optimize 
computational resources, 
we restrict the detailed calculations to 99 LS-coupled terms
rather than a 218-level fine structure expansion, and use a coarser energy 
mesh than \cite{n11} to demonstrate the importance of
enhanced background cross sections with broad PEC resonances and 
high-$n$ core excitations; lower resolution {\it per se} should not
lead to significant inaccuracy.

 \begin{figure}
 \begin{center}
\includegraphics[width=95mm,height=85mm]{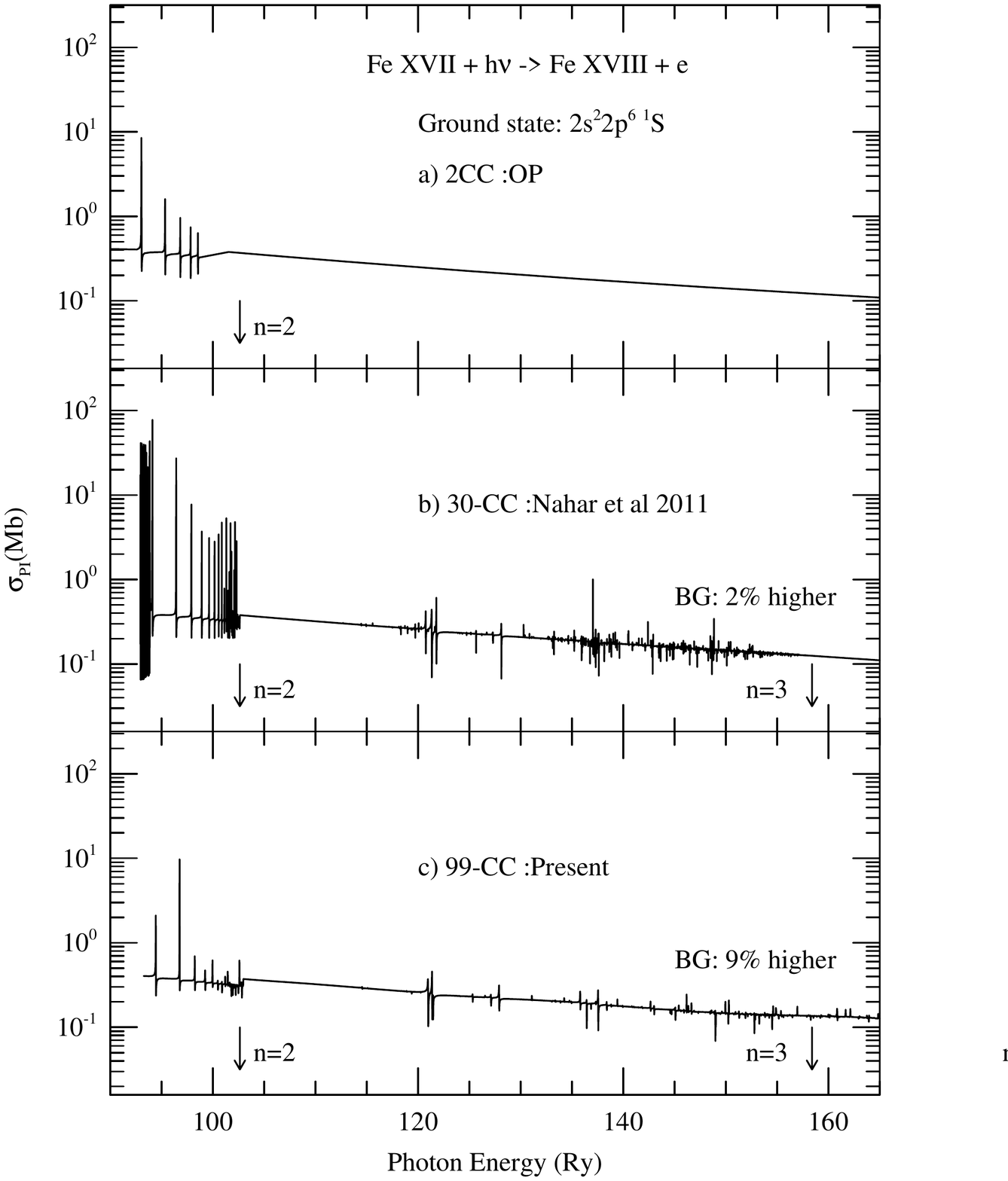}
 \end{center}
\vskip -0.35in
 \caption{Comparison of photoionization cross sections $\sigma_{PI}$ of the 
ground state of \fexvii $2s^22p^6(^1S)$, using three different wave function 
expansions for the core ion \fexviii: (a) 2CC, the Opacity Project 
(\cite{scott}, (b) 30CC (equivalent to 60 fine structure levels of \cite{n11}),
and (c) the present 99CC. Arrows point to the highest thresholds for \en = 2 
and 3 excitations (the \en = 4 thresholds up to 183.57 Ry are beyond the 
range shown). 'BG' is the background continuum at high energies; percentage 
enhancement relative to OP (a) is shown.
 \label{fig:gd}}
 \end{figure}
Fig. 2 compares the $\sigma_{PI}$ of the \fexvii ground state from three 
different calculations and core excitations: (a) OP --- 2CC, b) 30-CC, LS 
terms with $n\leq 3$ \cite{n11}, and (c) the present 99-CC, LS terms with 
$n\leq 4$. Except for differing resolution of resonances, all three 
$\sigma_{PI}$ have similar background cross sections; (b) and (c) are found 
to have small enhancements of 2\% and 9\% over the OP (a). Owing to fine 
structure splittings and higher resolution, the 30CC or 60-level cross 
sections (b), show more prominent resonances than the 99CC (c). 
 \begin{figure}
 \begin{center}
\includegraphics[width=95mm,height=95mm]{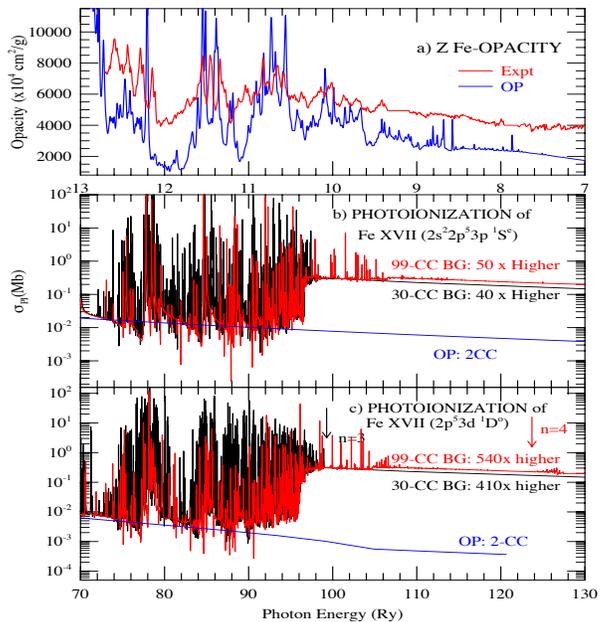}
 \end{center}
\vskip -0.35in
 \caption{Comparison of $\sigma_{PI}$ (panels b,c) of the excited $2s^22p^53p \ ^1S_0$ 
state of \fexvii from the three sets of core excitations: 2CC OP \cite{scott} 
(blue curves), 30CC \cite{n11}, and 99CC (present), with features in the 
opacity measurements at Sandia Z-pinch \cite{bail} (panel a). The energy 
range is the same, but in unit of \AA~for (a) and Ry for (b-c). Large 
enhancement factors relative to the OP (a) are marked.
 \label{fig:gd}}
 \end{figure}
In contrast to the ground state cross section, very large enhancements are 
found for excited state photoionization of \fexvii, exemplified in Figs.~3, 
4. Fig.~3a (top panel) show a detailed comparison of 
the OP monochromatic iron opacity with that measured at the highest 
temperature/ density achieved at the Sandia Z-machine, T = $2.11 \times 
10^{6}$ K and $n_e = 3.1 \times 10^{22}~cc$ \cite{bail}. 
The measured opacity 
spectra between 7-13\AA~corresponds to $\sigma_{PI}$ in 
lower panels in the energy range $\sim$ 70 - 130 Ry. The 
prominent transition arrays in the measurement are due to three dominant 
Fe ionization states: \fexvii, \fexviii and {\rm Fe~{\sc xix}}. 
The main points from the comparison are: (A) the measured 
background is consistently higher than OP throughout the energy range 
measured 7.5 $\leq \lambda$(\AA)$ \leq$ 12.5, (B) resonances are more broadened 
in experimental data, and (C) the ``windows" in theoretical OP opacity in 
between resonance complexes appear to be filled in with higher background 
opacity. 

Cross sections for the excited states $2s^22p^53p \ ^1S_0$ and 
$2s^22p^53d \ ^1D^o_2$ (Figs.~3b-c) are compared with the observed features 
in opacity (Figs.~3a). All three points (A,B,C) raised above can be
accounted for by the present results. 
Most revealing is the enhancement of the 
background (BG) cross sections, which is much higher than OP throughout the 
high energy range by up to factors of 40 and 50 at the last continuum energy 
for the 30CC $n \leq 3$ and the 99CC $n \leq 4$ expansions respectively
for the $\sigma_{PI} (2s^22p^53p \ ^1S_0)$ level (3b). 
An even larger background enhancement is seen for
$\sigma_PI (2s^22p^53d \ ^1D^o)$, up to factors of 
410 and 540 respectively compared to OP (3c). The enhancements are 
clearly related to the onset of the $n = 3$ and the $n = 4$ core excitation 
thresholds of \fexviii (c.f. Fig.~1). 
Furthermore, $\sigma_{PI}$ decreases only marginally, 
even up to high energies, in the entire energy range corresponding to the 
Z-measurements (3a). To estimate the enhancement quantitatively, we 
calculated the averaged oscillator strength $<f>_{\Delta E}$ (Eq.~3) using 
the $\sigma_{PI}$ (e.g. \cite{aas}) in the energy range of 70-130 Ry for the 
three states, the ground $2s^22p^6(^1S^e)$ and excited states 
$2s^22p^53p(^1S^e)$, $2s^22p^53d(^1D^o)$. 
The 2CC results with no excited $n$-complexes in the 
\fexviii expansion, underestimate the effective oscillator
strength by 
up to two orders of magnitude (with the exception of the ground state);
the $<f>$ values for the two excited states are  0.06, 1.81, 3.10 and
0.01, 1.86, 1.93 for 2CC, 30CC and 99CC calculations respectively.
Including $n$ = 4 core
levels raises the {\it background} even below the $n$ = 3 levels.
Resonances grow weaker, leading to 
a smooth background at the highest $n$ = 4 thresholds in 
Figs. 3b,c. Hence there is unlikely to be any significant enhancement by 
including higher $n > 4$ excitation thresholds (with far more cost), and 
the cross sections appear to have converged.
\begin{figure}
 \begin{center}
\includegraphics[width=95mm,height=70mm]{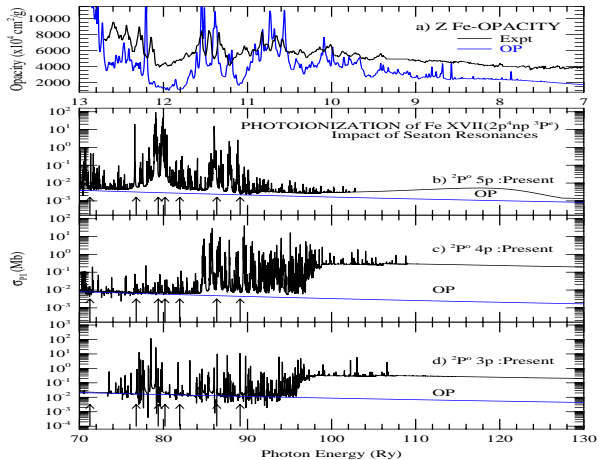}
 \end{center}
\vskip -0.35in
 \caption{Comparison of features in the measured opacity spectrum \cite{bail} (a),
with those due to PEC resonances in $\sigma_{PI}$ of a series of Rydberg 
states $2s^22p^5np (^2P^o) \ ^3P$ with \en = 3, 4 and 5 (b-d). Transition
energies corresponding to PEC resonances (not considered under OP 
\cite{scott}) are pointed out by arrows. The PECs are larger and wider  
resonances at higher excited states due to strong dipole core transitions 
that can enhance the background by orders of magnitude. The energy scale 
in panel (a) is in \AA~and in Ry in (b-d). 
 \label{fig:gd}}
 \end{figure}

Another prominent feature is autoionizing resonances due to strong dipole 
PEC transition arrays (2p \ra 3s,3d) $n\ell $ and 
(2p \ra 4s,4d) $n\ell$ in the core ion \fexviii, with the spectator electron 
$n\ell$. Such PEC or Seaton resonances (e.g. \cite{aas}) remain large in
photoionization of Rydberg levels with increasing $n$, even as the background 
decreases and $\sigma_{PI}$ starts at lower ionization potentials. Fig.~4 
shows a few of the computed $\sigma_{PI}$ of the Rydberg series 
levels: $2s^22p^5 (^2P^o)np \ ^3P$ with $n$ = 3,4,5. Whereas, the 
background enhancement is large for $n = 3p$ and $n = 4p$, it is much 
smaller for $n = 5$, while the resonances at PEC positions are higher. 
Photoionization of Rydberg levels is often taken to decrease with energy as 
a power-law. But for L-shell ions while it is approximately true at high 
energies as one approaches the K-edge, \cite{aas}, $\sigma_{PI}$ remains 
substantial as in the present calculations ({\it viz.} Fig.~3). Even for 
$\sigma_{PI}$ of a highly excited state $2s^22p^510p(^3P)$, the calculated 
$<f>$ is found to be much higher, 1.03, compared to 0.0013 from the OP cross 
sections. The energies of the strongest  dipole transitions 2p \ra 3s,3d and 
2p \ra 4s,4d arrays associated with the huge PEC resonances are marked with 
arrows in the Fig.~4. These transition arrays correspond to the inverse 
process of dielectronic recombination via satellite lines (2p \ra ns,nd) 
$n'\ell'$, with the spectator electron $n'\ell'$. 
Interestingly, some of 
the PEC resonances in \fexvii correspond to the energy region in between 
line (or resonant) features where the Z-opacity is much higher than the OP 
opacity (Fig-3a); including more ionization stages of Fe would further fill 
in the opacity. 

 Finally, we extend the atomic calculations to obtain monochromatic
\fexvii opacity at the Z plasma conditions shown in Fig.~5. The
measured opacity (5a) is higher than the OP (5d), as is the present
calculated opacity (5b,c). The
considerable structure obtained in the present 99CC calculations (Fig. 5c)
is mostly due to the large and broad PEC resonances, including
autoionization broadening in an {\it ab initio} manner. We carry out a
point-by-point normalized
Lorentzian convolution over all 454 \fexvii photoionization cross
sections using an algorithm that simulates temperature-density dependent
electron impact damping
\cite{bro}. It is seen that most resonances dissolve into and raise 
the continuum opacity, filling in the windows in the OP data, as seen
experimentally in 5a) \cite{bail}. That would also yield much
greater {\it continuum lowering} than existing opacity calculations.
The Rosseland Mean Opacity of \fexvii from the present results is 170
cm$^2$/g, 35\% higher than the 126 cm$^2$/g from OP. 
Substituting the present \fexvii opacity alone into a solar
mixture of all abundant elements from H to Ni, 
yields a 2.1\% increase. However, \fexvii is only one of the dominant
ions under
Z conditions with ionization fraction 0.195; \fexviii ionization fraction
is 0.39, twice as much. Given the generality of the bound-free
opacity enhancement shown in this {\it Letter}, including other contributing Fe
ions such as \fexviii and \fexix, would be consistent with 
the 7\% higher iron
opacity from {\it all} iron ions in the Z data (perhaps higher), 
and consequently with
the expected increase in total solar opacity that could solve the
abundances problem. Furthermore, the current approach treats the 
structures and divisions between the
bound-bound and bound-free opacities without unphysical
approximations.

Opacity calculations involve several other important atomic-plasma
issues.
First, plasma broadening significantly
affects the opacity distribution (c.f. Figs. 5b-c). Second, the Boltzmann
factors in the equation-of-state (EOS) for \fexvii imply low level
populations in excited states, but there are nearly 200 such levels
beginning 53 Ry above the ground state (Fig.~1) at the Z
temperature $\sim 2 \times 10^6$K ($\exp(-\Delta E/k_BT) =  0.015$), and
with fractional population between 0.1-1.0\% of the ground state;
augmented with orders of magnitude enhancement in cross sections 
that ensures a
significant contribution to high-energy bound-free opacity. Third, the
total oscillator strength sum-rule is sometimes invoked to rule out enhanced
opacity as measured \cite{iglesias}. However, it is the {\it partial},
not the total, differential  oscillator strength $\frac{df}{d\epsilon}$ 
in the relevant energy region, and atomic
species at a given temperature-density, that determines the mean opacity. 
There would be substantial re-distribution of
oscillator strength from the bound-bound, as in existing opacity models,
to the bound-free once the atomic/plasma effects demonstrated herein are
included. Fourth, the measured Fe ion fractions at the Z are close
to LTE values \cite{bail}. Considering the multitude of high-$n$ levels 
with large statistical weights \cite{pain}, there are 
marked differences in occupation
probabilities in the LTE EOS among different models \cite{tramp06}; sample
calculations show that OP values from the Mihalas-Hummer-D\"{a}ppen
\cite{mhd} EOS (also employed in this work) are {\it lower} than OPAL by 3\% 
for $n = 3$, a factor of 6 for $n = 5$, and by 2 orders of
magnitude for $n = 9$ \cite{bs03}; therefore, an upward revision would further
enhance the contribution of the present high-$n$ $\sigma_{PI}$ to
opacity. 
 \begin{figure}
 \begin{center}
\includegraphics[width=85mm,height=75mm]{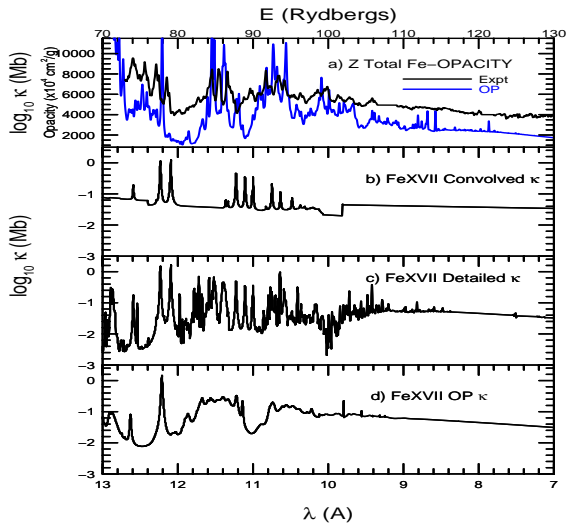}
 \end{center}
\vskip -0.35in
 \caption{a): Total iron opacity measured at Z and OP 
at T = 2.1 $\times 10^6$ K and $n_e = 3.1 \times 10^{22}$ cm$^{-3}$.
Present \fexvii opacity b)-c):
detailed autoionization resonance structures; c) 
convolved over electron impact broadening Lorentzian profile
\cite{bro}, which also fills in several windows in opacity
compared to OP results d).
Present values in b)-c) fall off more slowly than the OP d), and are
higher in the high-energy region towards 7 \AA.
 \label{fig:f6}}
 \end{figure}
The quantitative results presented herein help explain the observed 
discrepancies and the missing opacity due to photoabsorption from excited 
levels and core-excitations (the aforementioned points A, B and C).
The present work also points to a basic feature of photoionization of 
any atomic system: the bound-free cross sections would be incomplete unless
all contributing final states of the residual core ion are coupled.
However, that greatly enlarges the scope of photoionization calculations 
even on state-of-the-art 
computational platforms. Generally, the enhancement and convergence of 
photoionization cross sections up to high energies should manifest itself in 
the bound-free plasma opacity of many astrophysical and laboratory sources 
(e.g. \cite{drake}). 

We would like to thank Werner Eissner for contributions.
This work was partially supported by the U.S. Department of Energy
(DE-SC0012331) and and National Science Foundation (AST-1409207).  The 
computational work was carried out at the Ohio Supercomputer Center in 
Columbus, Ohio.

$\ast$ nahar.1@osu.edu, $\dagger$ pradhan.1@osu.edu


\begin{thebibliography}{10}
\bibitem{bro} A new algorithm has been developed for numerical 
simulation of temperature-density dependent 
electron impact damping of autoionizing
resonances in photoionization cross sections,
tabulated at thousands of energies for each of the 454 bound \fexvii
levels. Point-by-point convolution with a Lorentzian
functional is extremely CPU intensive, and would be described
elsewhere.
\end{thebibliography}
\end{document}